\documentclass[sigconf]{acmart}

\AtBeginDocument{%
  }

\setcopyright{none}  
\copyrightyear{2024}

\acmConference[arXiv]{arXiv}{April 2023}{Earth}

\renewcommand\footnotetextcopyrightpermission[1]{}

\usepackage{algorithmic}
\usepackage[linesnumbered,ruled,vlined]{algorithm2e}
\usepackage{graphicx}
\usepackage{textcomp}
\usepackage{xcolor}
\usepackage{color}
\usepackage{multirow}
\usepackage{subfigure}
\usepackage{CJKutf8}
\usepackage{wrapfig}
\usepackage{hyperref}
\usepackage{balance}
\usepackage{mathrsfs}
\usepackage{siunitx}
\usepackage{adjustbox}
\usepackage{booktabs}
\usepackage{bchart}
\usepackage{color,colortbl}
\usepackage{comment}
\usepackage{enumitem}
\usepackage{caption}
\usepackage{listings}
\usepackage[normalem]{ulem}

\usepackage{tcolorbox}

\usepackage{pifont}
\usepackage{booktabs}

\usepackage{xspace}
\usepackage{fontawesome}

\definecolor{Gray}{gray}{0.9}
\definecolor{darkgreen}{rgb}{0,0.5,0}

\usepackage{color,colortbl}
\usepackage{comment}

\usepackage{tikz}
\usepackage{pgfplots}

\usepackage{color,colortbl}
\usepackage{comment}
\definecolor{Gray}{gray}{0.9}
\definecolor{darkgreen}{rgb}{0,0.5,0}

\definecolor{gray91}{RGB}{91 128 184}
\definecolor{gray81}{RGB}{192 80 77}
\definecolor{gray71}{RGB}{245 195 66}
\definecolor{gray61}{RGB}{179 154 204}
\definecolor{gray51}{RGB}{186 218 84}

\usepackage{tikz}
\usepackage{pgfplots}
\usepackage{enumitem}

\usepackage{fontawesome}  

\usepackage{soul} 
\usepackage{color, xcolor}

\definecolor{bg}{HTML}{F8F9FB}  
\definecolor{bgc}{HTML}{FCF6E4}
\definecolor{rowcolor}{HTML}{ECEFF4}  
\definecolor{highlightY}{HTML}{FAE6A2}

\definecolor{mypurple}{HTML}{CEA8BC}
\definecolor{myblue}{HTML}{5E81AC}
\definecolor{mylightblue}{HTML}{81A1C1}
\definecolor{mysky}{HTML}{88C0D0}
\definecolor{mygreen}{HTML}{CBE0B8}

\usepackage[framemethod=TikZ]{mdframed}
\mdfdefinestyle{MyFrame}{%
    linecolor=black,
    outerlinewidth=0.3pt,
    roundcorner=5pt,
    skipabove = 5.5pt,
    skipbelow = 5.5pt,
    innertopmargin=5.5pt, 
    innerbottommargin=5.5pt, 
    innerrightmargin=5.5pt,
    innerleftmargin=5.5pt,
    backgroundcolor=bg, 
    }

\usepackage[frozencache,cachedir=./cache]{minted}
\definecolor{bblue}{HTML}{4F81BD}
\definecolor{rred}{HTML}{C0504D}
\definecolor{ggreen}{HTML}{9BBB59}
\definecolor{ppurple}{HTML}{9F4C7C}
\definecolor{nord15}{HTML}{B48EAD}
\definecolor{nordblue}{HTML}{EBCB8B}

\newcommand{\etal}{\hbox{\emph{et al.}}\xspace}
\newcommand{\eg}{\hbox{\emph{e.g.}}\xspace}
\newcommand{\ie}{\hbox{\emph{i.e.}}\xspace}

\begin{document}

\title{A study on Prompt Design, Advantages and Limitations of ChatGPT for Deep Learning Program Repair
  }

\author{Jialun Cao}
\affiliation{
  \institution{The Hong Kong University of Science and Technology, and
   Guangzhou HKUST Fok Ying Tung Research Institute}
  \country{China}
}

\author{Meiziniu Li}
\orcid{0000-0001-5947-4030}
\affiliation{%
  \institution{The Hong Kong University of Science and Technology
  }
  \country{China}
}

\author{Ming Wen}
\affiliation{%
  \institution{Huazhong University of Science and Technology}
  \country{China}
}

\author{Shing-Chi Cheung}
\affiliation{
  \institution{The Hong Kong University of Science and Technology, and
   Guangzhou HKUST Fok Ying Tung Research Institute}
  \country{China}
}

\renewcommand{\shortauthors}{CAO et al.}

\begin{abstract}
  ChatGPT has revolutionized many research and industrial fields. ChatGPT has shown great potential in software engineering to boost various traditional tasks such as program repair, code understanding, and code generation.
However, whether automatic program repair (APR) applies to deep learning (DL) programs is still unknown. DL programs, whose decision logic is not explicitly encoded in the source code, have posed unique challenges to APR. While to repair DL programs, an APR approach needs to not only parse the source code syntactically but also needs to understand the code intention. 
With the best prior work, the performance of fault localization is still far less than satisfactory (only about 30\%). Therefore, in this paper, we explore ChatGPT's capability for DL program repair by asking three research questions. (1) Can ChatGPT debug DL programs effectively? (2) How can ChatGPT's repair performance be improved by prompting? (3) In which way can dialogue help facilitate the repair? 
On top of that, we categorize the common aspects useful for prompt design for DL program repair.
Also, we propose various prompt templates to facilitate the performance and summarize the advantages and disadvantages of ChatGPT's abilities such as detecting bad code smell, code refactoring, and detecting API misuse/deprecation.
\end{abstract}





\maketitle

\section{Introduction}\label{sec:intro}

{Large language models (LLMs) such as ChatGPT have been receiving increasing attention due to their remarkable effectiveness in performing a wide range of tasks.~\footnote{\url{https://levelup.gitconnected.com/chatgpts-abilities-a-comprehensive-list-of-tasks-the-ai-language-model-can-perform-5a44492ab94}.}}
Recent advances in large language models (LLMs) have triggered various studies~\cite{codexLLM, llmEra, codeUnderstanding} examining the use of these models for software development tasks. 
{These studies analyze the capability of LLMs in program repair~\cite{codexLLM, llmEra}, code understanding, and code generation~\cite{codeUnderstanding}.} Their findings indicate that LLMs significantly outperform state-of-the-art techniques in program understanding~\cite{codeUnderstanding} and program repair~\cite{llmEra,codexLLM}. 
Furthermore, a recent study~\cite{sobania2023analysis} evaluates the program repair performance of ChatGPT on a benchmark consisting of 40 textbook Python programs (\eg, breadth-first-search, Hanoi, LCS-length).
The evaluation shows that ChatGPT achieves competitive performance
with state-of-the-art techniques. Furthermore, their preliminary attempt indicates that by taking advantage of prompting, the performance of ChatGPT can be further improved. 

In this paper, we study the dual perspectives (i.e., pros and cons) in the use of LLMs to debug and repair deep learning (DL) programs. We base our study on ChatGPT because of its impact, performance, and availability among comparable LLMs.
In addition, we put our focus on repairing DL programs for two main reasons. 
First, unlike conventional programs, a DL program manifests its functionality by guiding the training of the resulting deep neural network (DNN)~\cite{DeepFD}. This means the decision logic is not directly encoded in the source code. Thus, to repair the DL program, ChatGPT needs not only to parse the source code syntactically but also to understand the intention of DL programs to repair or optimize the resulting DNNs.
Besides the differences in paradigms, our study differs from prior studies~\cite{codexLLM, llmEra, codeUnderstanding, sobania2023analysis} in the selection of program subjects with significantly {more dependencies (6 times more libraries) and functionalities (more than the doubled number of lines). See Section~\ref{sec:benchmark} for more details.} 
Our study aims to answer three research questions.

\textbf{RQ1. Can ChatGPT debug the DL program effectively?}
This RQ is designed to understand ChatGPT's performance in DL program debugging as compared with that of two state-of-the-art techniques. In particular, we split the debugging task into three steps, \ie, fault detection, fault localization, and program repair. For evaluation, we quantify the number of faulty programs detected (fault detection ability), the number of faults localized (fault localization ability), and the number of faults repaired (program repair ability), respectively. As such, the pros and cons of ChatGPT could be better appreciated.

\textbf{RQ2. How can ChatGPT's repair performance be improved by prompting?}
As pointed out by several recent studies~\cite{nashidretrieval,sobania2023analysis,prompt-survey},
\textbf{\textit{prompt engineering}} can help LLMs provide better answers in various applications and research topics. 
{However, \textbf{\textit{what information should be included in the prompt}} remains unexplored. To answer this question, we turn to real-world queries raised by developers by analyzing and categorizing their questions. On top of our quantitative analysis, we propose an enhanced prompt template. Finally, we utilize this enhanced prompt template on the same benchmark to reveal its effectiveness.}

\textbf{RQ3. In which way can dialogue help facilitate the repair?} 
Since ChatGPT is designed to interact with users, whether and how performing dialogue could help to facilitate the program repair is unknown. Thus, we explore whether the hints of fault locations and incorrect repair could further improve the performance.

On top of that, we summarize the pros and cons of ChatGPT's ability during our study. To highlight the results, throughout the paper, we use three icons to annotate the pros ({\fontsize{10pt}{0}\selectfont\faThumbsOUp}), cons ({\fontsize{10pt}{0}\selectfont\faThumbsODown}), and features ({\fontsize{10pt}{0}\selectfont\faStar}) of utilizing ChatGPT in debugging DL programs.
For example, the deprecated API in DL libraries could affect the automatic DL program repair because ChatGPT fails to recognize the parameter names used in the old version. We also notice that, although the code refactoring could help simplify the code, some refactoring (\eg, variable renaming) could lead to program crashes due to inconsistent variable names. In addition, with the help of dialogue feature of ChatGPT, the program repair ability could be improved, yet meanwhile, with the input sequence increases, catastrophic forgetting or misunderstanding may happen to impede further improvements.

The contributions of this work are summarized as follows:
\begin{itemize}[leftmargin=*, topsep=1pt, itemsep=1pt]
    \item We conduct a study of the DL program debugging (\ie, fault detection, fault localization and program repair) ability of ChatGPT. The study complements prior works using buggy programs with more complicated functionalities and more dependencies.
    \item We categorize the common aspects useful for prompt design and propose enhanced prompt templates that are more effective for DL program repair.
    \item We summarize the characteristics of the programs repaired by ChatGPT. Our findings indicate that the program could be better repaired by providing more code intention. Also, ChatGPT is good at handling API misuse and deprecation, while it is short in default parameter identification and handling.
\end{itemize}
\section{Study Design}\label{sec:design}

\subsection{Benchmark}\label{sec:benchmark}
To evaluate the automatic DL program repair ability of ChatGPT, we use a benchmark released by a recent study~\cite{DeepFD}, which contains 58 buggy DL programs collected from StackOverflow and Github. 
In particular, 
each buggy DL program contains up to five types of faults, including inappropriate hyperparameter settings, incorrect data preprocessing, and incorrect model architectures. 
In the evaluation, we use the buggy programs collected from StackOverflow, because the programs in GitHub exceed the maximum input tokens.

Recent studies~\cite{refactoring,sobania2023analysis} explore the program repair ability of ChatGPT on conventional programs using different benchmarks~\cite{refactoring,QuixBugs}. 
Table~\ref{tab:benchmark-des} lists the differences between these benchmarks.
The two benchmarks consist of Python/Java programs implementing fundamental functionalities such as sorting algorithms.
Each program's average number of lines ranges from 13 to 22 and implements one functionality.
For comparison, in this paper, we use the benchmark~\cite{DeepFD} in which every program contains about 46 lines on average, implementing several functionalities including data preprocessing, DL model construction, model training, and evaluation. 
Furthermore, programs in the prior two benchmarks involve few dependencies, while the benchmark in this paper involves more than 6 dependencies for each program, including TensorFlow, Keras, and PyTorch. 
We hope our study could complement prior work by exploring ChatGPT's repair ability on more complicated programs and exploring the capabilities of handling dependency issues 
(\eg, software updates and dependency conflicts).

\subsection{Baselines}\label{sec:baseline}
In the study, we consider two baselines: AutoTrainer~\cite{autotrainer} and DeepFD~\cite{DeepFD}. AutoTrainer~\cite{autotrainer} identifies the training problems, such as dying reLU and oscillating loss during the training, and repairs the identified training problems using predefined repairing strategies. It iterates strategies until the training problem disappears or the maximum number of repairing is reached.
DeepFD~\cite{DeepFD} monitors the changes in runtime features (e.g., loss, gradient, and accuracy) during training and classifies the training process into predefined types of faults. Since DeepFD does not support program repair, we compare only its fault detection and localization abilities.

\subsection{ChatGPT Setup}
We build a pipeline for our experiments using the latest ChatGPT API based on the \texttt{gpt-3.5-turbo} model released by OpenAI.
Due to the non-deterministic nature of ChatGPT, for each query, we run five independent requests by starting a new conversation every time on the API. We record all prompts and their corresponding answers. We make them publicly available for review. 

\subsection{Evaluation Metrics}
We adopt the evaluation metrics following existing works~\cite{autotrainer,wardat21DeepLocalize,DeepFD,DeepDiagnosis}.
In the {fault detection} evaluation, we consider an alarm raised by an approach correct if it is made on a buggy program, regardless of whether the root cause has been identified. 
Due to the stochastic nature of ChatGPT, for each buggy program, we repeat the fault detection request five times, each with a new conversation.
If the majority ($\geq 3$) of replies correctly report fault existence, we denote the result as a \texttt{Yes}; otherwise, \texttt{No}. 
A fault localization is considered correct if the fault reported is the same as the one marked
  in the benchmark~\cite{DeepFD}. Since there could be more than one fault in a buggy program, we record the number of faults reported by each approach. For ChatGPT, we put down the number of faults reported each round, and we use the best result among the five rounds for comparison.
In the repair evaluation, a repair
is considered correct if the concerned fault is repaired in the same way as that in the benchmark. 
Note that we examine whether a repair is heading in the correct direction for the fault concerning incorrect training epochs. 
For example, for a buggy program, the training epoch is repaired from 1 to 50 in the benchmark, while ChatGPT repairs it to 20.
Although the repaired values of epochs are different, we consider it a correct repair since the repairing strategies (\ie, increasing the epoch) adopted are the same.

\begin{table}[t!]
    \centering
    \caption{Benchmark Comparison}\label{tab:benchmark-des}
    \renewcommand{\arraystretch}{1.3}
    \resizebox{1.0\linewidth}{!}{
    \begin{tabular}{l|l|c|c}
    \hline
    \textbf{Benchmarks} & \multicolumn{1}{c|}{\textbf{Description}} & \textbf{Ave Line} & \textbf{Ave Dep} \\
    \hline
    Refactory~\cite{refactoring} & \begin{tabular}[c]{@{}l@{}}Python programs regarding 5 basic tasks (e.g., \\sorting tuples, sequential search)\end{tabular} & 13.37 & 0.00 \\ 
    \hline
    QuixBug~\cite{QuixBugs} & \begin{tabular}[c]{@{}l@{}}Python and Java programs regarding 40 basic \\tasks (e.g., bitcount, breath-first-search). \end{tabular} & 22.32 & 0.44 \\
    \hline
    DeepFD~\cite{DeepFD} & \begin{tabular}[c]{@{}l@{}}DL programs in Python regarding \\various tasks (e.g., classification, regression)\end{tabular} & \textbf{45.91} & \textbf{6.70}\\
    \hline
    \end{tabular}
    }
    \end{table}

\section{RQ1. DL Program Debugging}
In this section, we explore ChatGPT's ability of DL program debugging against the benchmark. At first, we use the following \textbf{basic} prompt template (abbrev. \texttt{PT-B}) {adapted from a prior work~\cite{sobania2023analysis}}:


\begin{minted}[
    baselinestretch=1.0,
    fontsize=\footnotesize,
    xleftmargin=0.5ex,
    bgcolor=bg,
    breaklines=true,
    escapeinside=||
]{text}
|\underline{\textbf{Prompt Template (Basic)}}|: Is the following program buggy? If so, 
please repair it. |\sethlcolor{highlightY}\hl{[CODE]}|
\end{minted}

\noindent where \textbf{CODE} is filled in by every buggy DL program in the benchmark. The overall results for RQ1 are demonstrated in Table~\ref*{tab:rq1}. The results of ChatGPT using \texttt{PT-B} is denoted as (\texttt{ChatGPT$_\texttt{PT-B}$}) the table. 
In the following, we explain the results in detail.

\begin{table*}[t!]
    \centering
    \caption{\textbf{Comparison on Fault Detection, Fault Localization, and Program Repair}. (AT: AutoTrainer, DFD: DeepFD, ChatGPT$_{PT-B}$: ChatGPT with prompt template \texttt{PT-B} (RQ1), ChatGPT$_{PT-E}$: ChatGPT with prompt template \texttt{PT-E} (QR2)).
    }
    \label{tab:rq1}
    \renewcommand{\arraystretch}{1.2}
    \resizebox{1.0\textwidth}{!}{
    \begin{tabular}{lc|cccccc|cccccc|ccccc}
        \hline
        \textbf{} &  & \multicolumn{6}{c|}{\cellcolor[HTML]{CBE0B8}\textbf{Detection}} & \multicolumn{6}{c|}{\cellcolor[HTML]{D2E1F1}\textbf{Localization}} & \multicolumn{5}{c}{\cellcolor[HTML]{CEA8BC}\textbf{Repair}} \\
        \hline

         & \multicolumn{1}{c|}{} & \multicolumn{1}{c}{} & \multicolumn{1}{c}{} & \multicolumn{2}{c}{\cellcolor[HTML]{E5E9F0}} & \multicolumn{2}{c|}{\cellcolor[HTML]{FAE6A2}} & \multicolumn{1}{c}{\textbf{AT}} & \multicolumn{1}{c}{\textbf{DFD}} & \multicolumn{2}{c}{\cellcolor[HTML]{E5E9F0}\textbf{ChatGPT$_{PT-B}$}} & \multicolumn{2}{c|}{\cellcolor[HTML]{FAE6A2}\textbf{ChatGPT$_{PT-E}$}} & \multicolumn{1}{c}{\textbf{AT}} 
         & \multicolumn{2}{c}{\cellcolor[HTML]{E5E9F0}\textbf{ChatGPT$_{PT-B}$}} 
         & \multicolumn{2}{c}{\cellcolor[HTML]{FAE6A2}\textbf{ChatGPT$_{PT-E}$}} \\
        \multirow{-2}{*}{\textbf{Post}} & \multicolumn{1}{c|}{\multirow{-2}{*}{\textbf{\# F}}} & \multicolumn{1}{c}{\multirow{-2}{*}{\textbf{AT}}} & \multicolumn{1}{c}{\multirow{-2}{*}{\textbf{DFD}}} 
        & \multicolumn{2}{c}{\multirow{-2}{*}{\textbf{\begin{tabular}[c]{@{}c@{}}ChatGPT \\ {\cellcolor[HTML]{E5E9F0}${PT-B}$ }\end{tabular}}}} 
        & \multicolumn{2}{c|}{\multirow{-2}{*}{\textbf{\begin{tabular}[c]{@{}c@{}}ChatGPT \\ {\cellcolor[HTML]{FAE6A2}${PT-E}$}\end{tabular}}}} 
        & \textbf{Num} & \textbf{Num} 
        & {\cellcolor[HTML]{E5E9F0}\textbf{Num}} & {\cellcolor[HTML]{E5E9F0}\textbf{Best}} 
        & {\cellcolor[HTML]{FAE6A2}\textbf{Num}} & {\cellcolor[HTML]{FAE6A2}\textbf{Best}}
        & \textbf{Num} 
        & {\cellcolor[HTML]{E5E9F0}\textbf{Num}} & {\cellcolor[HTML]{E5E9F0}\textbf{Best}} 
        & {\cellcolor[HTML]{FAE6A2}\textbf{Num}} & {\cellcolor[HTML]{FAE6A2}\textbf{Best}} \\
        \hline
        48385830 & 3 & Y & Y & \textbf{0/5} & N & 5/5 & Y & 1 & 2 & 0/0/0/0/0 & 0 & 3/3/3/3/3 & 3 & 1 & 0/0/0/0/0 & 0 & \textbf{3/3/3/3/3} & 3 \\
        \rowcolor{rowcolor}
        44164749 & 4 & N & Y & 3/5 & Y & 5/5 & Y & 0 & 2 & 0/0/0/0/2 & 2 & 2/3/2/2/2 & 3 & 0 & 0/0/0/0/0 & 0 & 2/3/2/2/2 & 3 \\
        31556268 & 4 & N & Y & 3/5 & Y & 5/5 & Y & 0 & 3 & 4/0/0/0/3 & 4 & 3/3/1/3/2 & 3 & 0 & 1/0/0/0/0 & 1 & 1/1/0/1/1 & 1 \\
        \rowcolor{rowcolor}
        50306988 & 4 & Y & Y & 5/5 & Y & 5/5 & Y & 1 & 2 & 0/0/0/0/0 & 0 & 2/2/3/2/2 & 3 & 1 & 0/0/0/0/0 & 0 & 2/2/3/2/2 & 3 \\
        48251943 & 1 & N & Y & 4/5 & Y & 5/5 & Y & 0 & 0 & 0/1/1/0/1 & 1 & 1/1/1/1/1 & 1 & 0 & 0/\textbf{1/1}/0/0 & 1 & \textbf{1/1/1/1/1} & 1 \\
        \rowcolor{rowcolor}
        55328966 & 3 & Y & Y & 4/5 & Y & 5/5 & Y & 1 & 1 & 1/0/0/1/1 & 1 & 1/1/1/1/1 & 1 & 1 & 1/0/0/1/1 & 1 & 1/1/1/1/1 & 1 \\
        34311586 & 2 & N & Y & 5/5 & Y & 5/5 & Y & 0 & 2 & 1/1/1/1/1 & 1 & 1/1/1/1/1 & 1 & 0 & 1/1/1/1/1 & 1 & 1/1/1/1/1 & 1 \\
        \rowcolor{rowcolor}
        31880720 & 2 & N & Y & 5/5 & Y & 5/5 & Y & 0 & 2 & 2/1/1/1/1 & 2 & 2/2/1/2/2 & 2 & 0 & \textbf{2}/1/1/1/1 & 2 & \textbf{2/2}/1/\textbf{2/2} & 2 \\
        39525358 & 1 & N & Y & 5/5 & Y & 5/5 & Y & 0 & 0 & 0/0/0/0/0 & 0 & 0/0/0/0/0 & 0 & 0 & 0/0/0/0/0 & 0 & 0/0/0/0/0 & 0 \\
        \rowcolor{rowcolor}
        48934338 & 1 & N & Y & 5/5 & Y & 5/5 & Y & 0 & 0 & 0/0/0/0/0 & 0 & 1/1/1/0/1 & 1 & 0 & 0/0/0/0/0 & 0 & \textbf{1/1/1}/0/\textbf{1} & 1 \\
        59325381 & 1 & N & Y & 5/5 & Y & 5/5 & Y & 0 & 1 & 1/1/1/1/1 & 1 & 1/1/1/1/1 & 1 & 0 & \textbf{1/1/1/1/1} & 1 & \textbf{1/1}/0/\textbf{1/1} & 1 \\
        \rowcolor{rowcolor}
        59278771 & 3 & N & Y & 4/5 & Y & 5/5 & Y & 0 & 1 & 0/0/0/0/0 & 0 & 1/1/1/1/1 & 1 & 0 & 0/0/0/0/0 & 0 & 1/1/1/1/1 & 1 \\
        48221692 & 1 & N & Y & 5/5 & Y & 5/5 & Y & 0 & 0 & 0/0/0/0/0 & 0 & 1/1/1/1/1 & 1 & 0 & 0/0/0/0/0 & 0 & \textbf{1/1/1/1/1} & 1 \\
        \rowcolor{rowcolor}
        50079585 & 5 & Y & Y & 2/5 & N & 5/5 & Y & 1 & 2 & 0/0/0/0/0 & 0 & 3/3/3/3/3 & 3 & 1 & 0/0/0/0/0 & 0 & 2/2/2/2/2 & 2 \\
        45337371 & 1 & N & Y & 5/5 & Y & 5/5 & Y & 0 & 1 & 1/1/1/1/1 & 1 & 1/1/1/1/1 & 1 & 0 & 0/\textbf{1/1/1}/0 & 1 & 0/0/0/0/0 & 0 \\
        \rowcolor{rowcolor}
        44066044 & 2 & N & Y & 5/5 & Y & 5/5 & Y & 0 & 0 & 1/1/1/0/1 & 1 & 2/2/1/2/2 & 2 & 0 & 1/1/1/0/1 & 1 & \textbf{2/2}/1/\textbf{2/2} & 2 \\
        47352366 & 1 & Y & Y & 5/5 & Y & 5/5 & Y & 1 & 1 & 1/1/1/1/1 & 1 & 1/1/1/1/1 & 1 & 1 & \textbf{1/1/1/1/1} & 1 & \textbf{1/1/1/1/1} & 1 \\
        \rowcolor{rowcolor}
        45442843 & 4 & N & Y & 4/5 & Y & 5/5 & Y & 0 & 2 & 1/0/1/0/1 & 1 & 3/3/2/4/4 & 4 & 0 & 0/0/0/0/0 & 0 & 2/2/1/\textbf{4/4} & 4 \\
        48594888 & 2 & N & Y & 2/5 & N & 5/5 & Y & 0 & 1 & 0/0/0/0/0 & 0 & 1/2/2/2/2 & 2 & 0 & 0/0/0/0/0 & 0 & 1/\textbf{2/2/2/2} & 2 \\
        \rowcolor{rowcolor}
        31627380 & 2 & N & Y & 3/5 & Y & 5/5 & Y & 0 & 1 & 0/0/0/0/0 & 0 & 0/0/0/0/0 & 0 & 0 & 0/0/0/0/0 & 0 & 0/0/0/0/0 & 0 \\
        58609115 & 1 & N & Y & 5/5 & Y & 4/5 & Y & 0 & 0 & 0/1/1/0/0 & 1 & 1/1/0/1/1 & 1 & 0 & 0/0/0/0/0 & 0 & 0/\textbf{1}/0/0/0 & 1 \\
        \rowcolor{rowcolor}
        50481178 & 2 & N & Y & 1/5 & N & 5/5 & Y & 0 & 0 & 0/0/0/0/0 & 0 & 2/2/1/1/1 & 2 & 0 & 0/0/0/0/0 & 0 & 1/1/1/1/1 & 1 \\
        56380303 & 2 & N & Y & 4/5 & Y & 5/5 & Y & 0 & 0 & 0/0/0/0/0 & 0 & 2/2/2/2/2 & 2 & 0 & 0/0/0/0/0 & 0 & \textbf{2/2/2/1/2} & 2 \\
        \rowcolor{rowcolor}
        58237726 & 2 & Y & Y & 1/5 & N & 5/5 & Y & 0 & 1 & 0/0/0/0/0 & 0 & 1/1/1/1/1 & 1 & 0 & 0/0/0/0/0 & 0 & 0/0/1/0/0 & 1 \\
        37624102 & 1 & Y & Y & 5/5 & Y & 5/5 & Y & 1 & 1 & 0/0/0/0/0 & 0 & 1/1/1/1/1 & 1 & 1 & 0/0/0/0/0 & 0 & \textbf{1/1/1/1/1} & 1 \\
        \rowcolor{rowcolor}
        39810655 & 1 & N & Y & \textbf{0/5} & N & 5/5 & Y & 0 & 1 & 0/0/0/0/0 & 0 & 0/0/0/0/0 & 0 & 0 & 0/0/0/0/0 & 0 & 0/0/0/0/0 & 0 \\
        41600519 & 4 & Y & Y & 5/5 & Y & 5/5 & Y & 1 & 1 & 2/1/1/2/2 & 2 & 1/2/2/2/2 & 2 & 1 & 1/0/1/1/2 & 2 & 1/2/2/2/2 & 2 \\
        \rowcolor{rowcolor}
        41977498 & 3 & N & Y & 5/5 & Y & 5/5 & Y & 0 & 0 & 0/0/1/0/0 & 1 & 0/1/1/0/0 & 1 & 0 & 0/0/1/0/0 & 1 & 0/1/1/0/0 & 1 \\
        42800203 & 1 & N & Y & 5/5 & Y & 5/5 & Y & 0 & 0 & 1/1/0/1/1 & 1 & 1/1/1/1/1 & 1 & 0 & \textbf{1/1}/0/\textbf{1/1} & 1 & \textbf{1/1/1/1/1} & 1 \\
        \rowcolor{rowcolor}
        44998910 & 1 & N & Y & 3/5 & Y & 5/5 & Y & 0 & 0 & 0/0/0/0/0 & 0 & 0/0/0/0/0 & 0 & 0 & 0/0/0/0/0 & 0 & 0/0/0/0/0 & 0 \\
        46995209 & 4 & N & Y & 5/5 & Y & 5/5 & Y & 0 & 0 & 2/2/2/1/1 & 2 & 2/1/1/2/2 & 2 & 0 & 2/2/2/1/1 & 2 & 1/0/0/1/1 & 1 \\
        \rowcolor{rowcolor}
        51181393 & 1 & N & Y & 1/5 & N & 5/5 & Y & 0 & 1 & 0/0/0/0/0 & 0 & 0/1/1/1/1 & 1 & 0 & 0/0/0/0/0 & 0 & 0/\textbf{1/1/1/1} & 1 \\
        47932589 & 1 & N & Y & 5/5 & Y & 5/5 & Y & 0 & 0 & 0/0/0/0/0 & 0 & 1/1/1/1/0 & 1 & 0 & 0/0/0/0/0 & 0 & 0/0/0/0/0 & 0 \\
        \rowcolor{rowcolor}
        37213388 & 1 & N & Y & 5/5 & Y & 5/5 & Y & 0 & 0 & 0/0/0/0/0 & 0 & 1/1/1/1/1 & 1 & 0 & 0/0/0/0/0 & 0 & \textbf{1/1/1/1/1} & 1 \\
        \hline
        & 72 & 8 & \textbf{34} &  & 27 &  & \textbf{34} & 7 & 29 &  & 23 &  & \textbf{50} & 7 &  & 16 &  & \textbf{43}\\
        \bottomrule
    \end{tabular}
    }
    \end{table*}

\subsection{Fault Detection}
We first examine whether the approaches can detect existing faults. The results are demonstrated in Table~\ref{tab:rq1} (\textbf{Detection}).
In particular, for each buggy program, we report whether the approaches can detect the existence of faults or not, which are labeled as \texttt{Y/N} (\texttt{Yes/No}). 
For ChatGPT, we report five times of \texttt{Y/N}.

We can see from the table that the baseline DeepFD (DFD) can detect all buggy DL programs (34/34), outperforming \texttt{ChatGPT$_\texttt{PT-B}$}, which detects 27 of the 34 buggy programs. There are seven buggy programs where ChatGPT fails to detect faults in the majority of five rounds. For two (\ie, 48385830 and 39810655) of the seven, no faults can be detected by ChatGPT in all five rounds. 
An example is shown below:
\begin{minted}[
    baselinestretch=1.3,
    fontsize=\footnotesize,
    xleftmargin=0.5ex,
    bgcolor=bg,
    breaklines=true,
    escapeinside=||
]{text}
|\underline{\textbf{Answer}}:| The program appears to be |\textbf{functioning correctly}| and there 
are |\textbf{no obvious bugs}|.
\end{minted}

\noindent 
For four (\ie, 50079585, 50481178, 58237726, and 51181393) of the seven, ChatGPT reports the existence of faults in at least one round and attempts to ask for more information, such as the intention of the code and the data used for training, in some other rounds. An example of ChatGPT asking for more information is shown below. 

\begin{minted}[
    baselinestretch=1.3,
    fontsize=\footnotesize,
    xleftmargin=0.5ex,
    bgcolor=bg,
    breaklines=true,
    escapeinside=||
]{text}
|\underline{\textbf{Answer}}:| The program |\textbf{does not have any syntax errors or logical}| 
|\textbf{errors}| that are immediately apparent. However, it is difficult to 
determine if it is fully functional |\underline{\textit{\textbf{without knowing the context and}}}| 
|\underline{\textit{\textbf{intention of the code.}}}|
\end{minted}

\noindent The reply indicates the need for more informative prompts.

\begin{mdframed}[style=MyFrame]
    \textbf{Finding 1}: 
    Using \texttt{PT-B} as the prompt template, 
    ChatGPT can detect 27/34 buggy DL programs, while the DeepFD baseline can detect all of them (34/34). 
    Some replies from ChatGPT suggest that \textit{\textbf{more information (\eg, the context and intention of code) is needed in the prompt}}.
    \end{mdframed}

DL program faults detected by ChatGPT broadly fall into three types: syntactic faults, functional faults, and code smell. In the following, we discuss how ChatGPT provides answers for these fault types. 
(1) \textbf{Syntactic Faults}. ChatGPT usually 
reports whether a program is syntactically correct at the beginning of the reply. A program that contains a syntactic fault (\eg, undefined variables, missing importing dependencies) can prevent ChatGPT from detecting the other types of faults. 
(2) \textbf{Functional Faults}. This type of fault usually causes the program fails to perform its intended functions correctly. To detect such faults, ChatGPT needs to understand the code intention, then detect and repair the fault \textit{when the} \textit{implementation differs from the code intention} as understood by {ChatGPT}. 
For example, if a DL program first processes the labels using \texttt{one-hot} encoding (\ie, the pre-process indicates that it is a dataset for classification problem), then uses \texttt{mean\_absolute\_error} (which is inappropriate for the classification problem) to calculate the loss.
In such cases, there is a gap between the implementation and the functionality as ChatGPT understands, so it reports a functionality fault. 
Another interesting finding is that ChatGPT has learned to detect missing or redundant implementation. For example, it suggests adding codes to visualize the training process or removing the redundant print statement.
(3) \textbf{Code Smell}. Code smells do not necessarily result in functional errors. They are more likely poor coding practices. ChatGPT is able to suggest possible ways to improve the program's quality (\eg, code refactoring) or performance (\eg, replacing self-implemented methods with the available APIs). 

\begin{mdframed}[style=MyFrame]
    \textbf{Finding 2}: 
    {\fontsize{8pt}{0}\selectfont \faStar}
    Three types of faults that are observed could be detected by ChatGPT, including \textit{syntactic faults}, \textit{functional faults}, and \textit{code smells}.
    {\fontsize{10pt}{0}\selectfont\faThumbsOUp}
     We highlight that \textit{\textbf{the detection of functional fault involves the understanding of the code intention}}. 
     In other words, the detection ability of these faults could be further improved when code intention is introduced.
    \end{mdframed}

\subsection{Fault Localization}
Accurate fault localization is a critical step toward automatic program repair, so we check whether the faults in the program can be correctly localized. Since there could be more than one fault in one program, we show the number of faults that are correctly localized by each approach. For ChatGPT, we list the results of each round and highlight the best results over five rounds. 

As shown in Table~\ref*{tab:rq1}, the best existing work can localize 29/72 faults correctly. ChatGPT (PT-B) localizes slightly fewer faults, with 23/72. 
The results are far less than satisfactory. 
Via further analysis, we found that among the 49 (72 - 23) escaped faults that belong to 18 cases, 16 of them (belonging to 7 cases) cannot be even detected, 32 of them (belonging to 11 cases) can be detected but cannot be correctly-localized. 
In particular, we take a closer analysis of these 32 faults in 11 cases, and we observe that instead of identifying faults that cause severe consequences (\eg, missing data preprocess),
ChatGPT tends to localize the \textbf{outdated APIs} (39525358, 59278771, and 37624102), the \textbf{bad code smells} (44998910, 48221692, 50306988, 48934338, 31627380, 37213388, 47932589)
when it is asked to repair the program without explicit repair goal (\ie, improve the accuracy) provided.
Although the above improvements could benefit the programs' quality, they are not necessarily serious faults.  
For example, adding more comments does improve the code readability, while the more serious faults (\eg, missing data preprocessing, improper training strategy) are worth more attention yet are escaped from the fault localization.

\begin{mdframed}[style=MyFrame]
    \textbf{Finding 3}:   
    {\fontsize{8pt}{0}\selectfont \faStar}
    When asked to repair the program without an explicit repair goal,
    ChatGPT tends to focus on identifying \textit{\textbf{outdated APIs}} and \textit{\textbf{bad code smells}}.
    \end{mdframed}

Furthermore, we observe that the missing or incorrect localization is mainly due to the misunderstanding of the code intention.
For example, in one case (56380303), ChatGPT falsely localizes the activation function in the output layer as a fault, because it mistakenly believes the DL program is trained for regression program. While it turned out to be a classification problem, thus the localized fault is a false alarm. 
Another example~\cite{so48221692}
shows a missing fault localization due to an incomplete understanding of the desired task:

\begin{minted}[
    baselinestretch=1.0,
    fontsize=\scriptsize,
    xleftmargin=0.5ex,
    bgcolor=bg,
    breaklines=true,
    escapeinside=||
]{python}
|\underline{\textbf{Prompt}}:|
    dataset = pd.read_csv("simple_network/Linear Data.csv", 
                           header = None).values
    X_train, X_test, Y_train, Y_test = train_test_split(dataset[:,0:1], 
                                                        dataset[:,1], 
                                                        test_size=0.25,)
    # Now we build the model
    neural_network = Sequential() # create model
    neural_network.add(Dense(5, input_dim=1, activation='sigmoid')) 
    neural_network.add(Dense(1, activation='sigmoid')) # output layer
                     
|\underline{\textbf{Answer}}:| The program seems to be correct, but we can make some minor improve-
ments. First, we can remove the commented out lines, as they are not necessary.
Second, we can add some comments to explain what is happening. Finally, we 
can print the final loss and accuracy of the neural network on the test set.
\end{minted}

\noindent For the above code, ChatGPT localizes the faults regarding the bad code smells (\eg, redundant comments). However, the DL program is trained on a linear dataset,
and the developer found that all the outputs range from 0 to 1. The fault is caused by the improper activation function \texttt{sigmoid}, whose output can only range from 0 to 1. Without knowing the specific learning task, ChatGPT can hardly find out the fault.

\begin{mdframed}[style=MyFrame]
    \textbf{Finding 4}: 
    Knowing \textit{\textbf{the learning task}} and \textit{\textbf{the objective of the repair}} could be useful for better fault localization. It suggests involving such information in the prompt.
    \end{mdframed}

\subsection{Program Repair}
Given the performance on fault localization, the unsatisfactory performance on program repair is reasonable. 
From Table~\ref{tab:rq1} \texttt{ChatGPT$^{PT-B}$}, we can see that only 16/72 faults can be correctly repaired by ChatGPT. This number doubles that achieved by the baseline (7/72). In addition, it is noticeable that ChatGPT can repair 6 cases (48251943, 31880720, 59325381, 45337371, 47352366, 42800203) completely, while AutoTrainer can only repair two cases (47352366, 37624102) completely.

Furthermore, among the replies, we summarize the ways ChatGPT replies to the repaired program. There are three main ways of replying. (1) \textbf{Explanation with the complete repaired code}. 
This way is straightforward and usually comes with an explanation regarding the repaired strategies. 
One could directly execute the program with the complete code to see whether the repair works.
An example is shown below~\cite{so33969059}:

\begin{minted}[
    baselinestretch=1.0,
    fontsize=\footnotesize,
    xleftmargin=1.5ex,
    bgcolor=bg,
    breaklines=true,
    escapeinside=||
]{text}            
|\underline{\textbf{Answer}}:| Yes, there is a bug in the program. The activation 
function for the output layer should be a sigmoid instead of 
softmax because the problem is binary classification. 
The repaired program is: (Code is omitted)
\end{minted}

\noindent However, it is noteworthy that code refactoring happens commonly. For example, the variable \texttt{X} and \texttt{Y} are renamed as \texttt{x\_train} and \texttt{y\_train} in some cases (\eg, 44066044). 

(2) \textbf{Code snippets with explanations}. This kind of reply does not contain the complete repaired code. Instead, it contains only code snippets regarding the repair. An example is provided~\cite{so37624102}:

\begin{minted}[
    baselinestretch=1.0,
    fontsize=\scriptsize,
    xleftmargin=0.5ex,
    bgcolor=bg,
    breaklines=true,
    escapeinside=||
]{python}
|\underline{\textbf{Answer}}:| The program is not buggy. However, there is a warning that the 
"nb_epoch" parameter in the "fit" function is deprecated and will be removed 
in a future version of Keras. It is recommended to use the "epochs" parameter 
instead. So, to prevent future errors, replace "nb_epoch" with "epochs" as 
follows:
    model.fit(X_train, Y_train, batch_size=32, epochs=10)
\end{minted}

\noindent Such replies could be mechanically used for repairing by replacing the buggy lines with the repaired ones. However, there are some cases where the returned code snippet renames the variables. Directly replacing such code lines could lead to a crash.

(3) \textbf{Explanation of the repair strategies}. In these replies, there is no or little code provided. ChatGPT makes suggestions on the repair strategy, while usually, the concrete fixes are not clear. In such cases, users need to repair the code manually according to the suggestions, and some try-out is needed to find the optimal fixes. In the following, we show an example reply returned by ChatGPT for a case in the benchmark~\cite{so31556268}
: 

\begin{minted}[
    baselinestretch=1.0,
    fontsize=\scriptsize,
    xleftmargin=0.5ex,
    bgcolor=bg,
    breaklines=true,
    escapeinside=||
]{text}
|\underline{\textbf{Answer}}:| To improve the program, some possible modifications can be: (1) 
Changing the activation function to ReLU or other non-saturating functions, 
and experiment with different architectures. (2) Using a learning rate 
scheduler or adaptive optimizers like Adam to improve convergence. 
(3, 4, and 5 are omitted due to space limitation.)
\end{minted}

\noindent In this case, five suggestions are made to repair the program. 
These suggestions provide a rough direction about how to repair (\eg, experiment with different architectures and hyperparameters). The concrete fixes still need to be found out by developers. This kind of reply cannot be directly executed, and rely on manual checking and multiple rounds of attempts.

\begin{mdframed}[style=MyFrame]
    \textbf{Finding 5}: 
    {\fontsize{8pt}{0}\selectfont \faStar}
    There are three main types of ChatGPT's response for repair, including the \textit{Explanation with the complete repaired code}, \textit{Code snippets with explanation}, and \textit{Explanation of the repair strategies}. The first two ways can be validated by execution with no or little effort. 
    {\fontsize{10pt}{0}\selectfont\faThumbsODown}
    While one should be careful with \textit{\textbf{code refactoring}} and \textit{\textbf{variable renaming}}, in case they cause a crash. The third way relies on manual effort, which may not be an optimal solution to automatic repair. One can avoid it by explicitly asking ChatGPT to return the complete repaired code.
    \end{mdframed}

\section{RQ2. Impact of Prompt}\label{sec:rq2}
The experimental results of ChatGPT in RQ1 are less than satisfactory, we are then motivated to design a better prompt to improve the program repair performance utilizing the power of ChatGPT.

\subsection{Study of Real-world Questions}

As revealed in RQ1, providing more information, such as code intention, is desired for prompt design.
Yet, \textbf{\textit{what information should be included in the prompt?}}
To better explore this question, we turn to the queries raised by real developers. 
In particular, we scraped the texts (code blocks are excluded) of those cases from StackOverflow, and then carefully categorized the descriptions into seven types of information as follows via manual analysis. 

\textbf{\textit{Symptom}} clarifies the user intention towards repairing. It allows ChatGPT to better understand the symptom that users want to repair (\eg, syntax, dependency, or performance error). The common symptoms include  abnormal outputs (\eg, NaN, zero output), unsatisfactory results (\eg, low accuracy, huge loss), over-fitting, or slow convergence. 

\textbf{\textit{Task}} explains the design purpose of the code. Both loss functions \texttt{categorical\_crossentropy} and \texttt{mean\_absolute\_} \texttt{error} are syntactically correct in DL programs. While given the context that the program is designed for classification problems, the first loss function is more desirable than the latter one.
The common tasks include classification problems (\eg, multi-label or binary classification) and regression problems.

\textbf{\textit{Dataset}} describes the characteristics of the input data (features and labels). Obtaining the dataset's knowledge could help ChatGPT to gain a better understanding of the code intention. For example, if ChatGPT knows the input data are images, it has a higher chance of identifying the proper data preprocessing method, loss function, etc. In addition, input data range also affects the way of data preprocessing. For example, knowing the input values ranging from 0 to 255, it is more likely that ChatGPT could point out a normalization is needed before training. Generally, the description regarding the dataset includes the type of training data (\eg, images), the distribution of data (\eg, a sequence of sine waves), positive/negative values, the dimension of training data, and the number of labels. It is also the case where the dataset is well-known, such as MNIST and Iris. In that case, specifying the dataset name is informative and helpful.

\textbf{\textit{Version/Dependency}} specifies the version of libraries that are used. Since the libraries are evolving rapidly, specifying the version of libraries may help to reproduce the code correctly and to get the proper API recommendation. The versions of Python, the versions of DL libraries (\eg, PyTorch, Tensorflow), and the version of the backends are commonly seen in StackOverflow.

\textbf{\textit{Attempted fixes}} are those fixes that have been tried by Developers but failed to achieve the desired goals. The attempted fixes help to avoid ineffective repair strategies.

\textbf{\textit{Hyperparameter}} and \textbf{\textit{Model Architecture}} provide a brief introduction to the network and the training strategies. While this information could be easily identified in code, cases regarding such information are rare.

\begin{figure}[t!]
    \centering
    \setlength{\belowcaptionskip}{-15pt}
    \resizebox{0.92\linewidth}{!}{
    \includegraphics{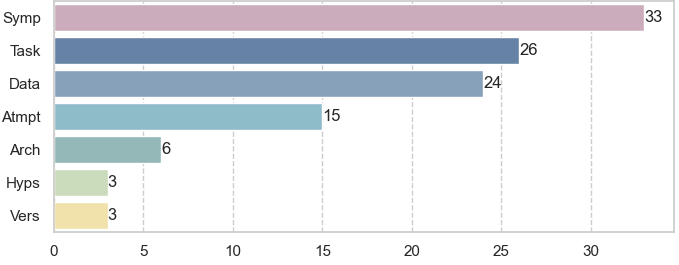}
    }
\caption{\textbf{Statistics of Question Descriptions.}}\label{fig:questions}
\end{figure}

We visualize the statistics of each type of information in Figure~\ref{fig:questions}. Clear that the description of \textit{symptom} is observed in almost all cases (33/34). Followed by the \textit{task} (26/34) and \textit{dataset} (24/34) descriptions. 
The remaining information is observed occasionally.

\begin{mdframed}[style=MyFrame]
    \textbf{Finding 6}: 
    {\fontsize{10pt}{0}\selectfont\faLightbulbO}
    There are mainly seven types of query information that are commonly seen in real-world forums. Among them, the description regarding \textit{\textbf{symptom}, \textbf{task}, and \textbf{dataset}} are the most commonly observed ones. 
    \end{mdframed}

\subsection{Enhanced Prompt Template Design}\label{sec:template}
On top of the observation from real-world questions, we propose an enhanced prompt template (abbrev. \texttt{PT-E}) as follows:

\begin{minted}[
    baselinestretch=1.0,
    fontsize=\footnotesize,
    xleftmargin=0.5ex,
    bgcolor=bg,
    breaklines=true,
    escapeinside=||
]{text}
|\underline{\textbf{Prompt Template (Enhanced)}}:| The following code is designed for a
|\sethlcolor{highlightY}\hl{[X]}| problem trained on |\sethlcolor{highlightY}\hl{[Y]}|.Please repair it in order to |\sethlcolor{highlightY}\hl{[Z]}|. [CODE]
\end{minted}

\noindent where \textbf{X} describes the task (\eg, classification, regression), \textbf{Y} specifies the dataset (\eg, MNIST, ImageNet, numbers that follow the sine waves), and \textbf{Z} describes the objective (\eg, improve the accuracy) against the symptom. An example prompt is shown below:

\begin{minted}[
    baselinestretch=1.0,
    fontsize=\footnotesize,
    xleftmargin=0.5ex,
    bgcolor=bg,
    breaklines=true,
    escapeinside=||
]{text}
|\underline{\textbf{Prompt Template (Enhanced)}}:| The following code is designed for 
a |\underline{classification}| problem trained on |\underline{Iris dataset}|. Please repair it in 
order to |\underline{improve the accuracy}|. [CODE]
\end{minted}

\noindent In case any of the X/Y/Z is unavailable, we adjust the prompt slightly as follows. 

\begin{minted}[
    baselinestretch=1.0,
    fontsize=\footnotesize,
    xleftmargin=0.5ex,
    bgcolor=bg,
    breaklines=true,
    escapeinside=||
]{text}
|\underline{\textbf{Prompt Template (when X is absent)}}:| The following code is trained on 
|\sethlcolor{highlightY}\hl{[Y]}|. Please repair it in order to |\sethlcolor{highlightY}\hl{[Z]}|. 
|\underline{\textbf{Prompt Template (when Y is absent)}}:| The following code is designed 
for a |\sethlcolor{highlightY}\hl{[X]}| problem. Please repair it in order to |\sethlcolor{highlightY}\hl{[Z]}|. 
|\underline{\textbf{Prompt Template (when Z is absent)}}:| The following code is designed 
for a |\sethlcolor{highlightY}\hl{[X]}| problem. Please repair it. 
\end{minted}

\subsection{Effectiveness of the Enhanced Prompt}
We use the above enhanced prompt template to rerun the experiment against the benchmark, examining whether the effectiveness of fault detection, localization, and repair could be enhanced. Note that for the cases where the learning task/dataset/objective is unavailable, we use the variant templates as listed in Section~\ref{sec:template}. 

The experimental results are shown in Table~\ref{tab:rq1} (\texttt{ChatGPT$_\texttt{PT-E}$}). 
We can see that the improvement is significant. For fault detection, all the buggy programs can be detected, achieving a comparable result to the best existing approach (34/34), compared with 27/34 achieved when using the basic prompt template. 
Besides, the confidence of \texttt{ChatGPT$_\texttt{PT-E}$} is much higher than \texttt{ChatGPT$_\texttt{PT-B}$}, \ie, ChatGPT with enhanced prompt reports the fault exists in almost every round (5/5). 
For fault localization, the number of correctly localized faults doubles from 23 to 50. 
The number of correctly repaired faults triples from 16 to 43. 
In addition, with the aid of the enhanced prompt template, there are 16/34 cases that can be completed and repaired, compared with only 6/34 using the basic template. 
The significant improvement shows the effectiveness of the enhanced prompt template. 

We also notice that after providing the target of repair (\ie, \texttt{[Z]} in the template) to ChatGPT, the number of cases where bad code smells and extra functionalities are repaired decreases, while the true positives increases. It indicates that 
with intentions, the repair could be more accurate.

\begin{mdframed}[style=MyFrame]
    \textbf{Finding 7}: 
    {\fontsize{8pt}{0}\selectfont\faThumbsOUp ~\faStar}
    The enhanced prompt template helps to clarify the intention regarding the purpose of the {code design, the symptom that needs to be alleviated, and the dataset characteristics}. \textit{\textbf{With clearer intention introduced, the focus of repair shifted from code smells to functionality faults}}.
    \end{mdframed}


\section{RQ3. Effectiveness of ChatGPT's Dialogue}
One important feature of ChatGPT is user interaction, so we further explore whether the DL program repair ability can be enhanced by providing more information via dialogues. As demonstrated in a recent study~\cite{sobania2023analysis}, the dialogue can improve the final repair for conventional program repair.
However, \textit{\textbf{how to carry out multi-round dialogues with ChatGPT {in DL program repair} effectively} remains unknown.} We then push the study forward, seeking better ways to exploit the dialogue feature.

First, we involve the fault location information using the following prompt template.  
In particular, for the cases that failed to be completely repaired 
in Section~\ref{sec:rq2} (\ie, 18 cases), we first provide the \textbf{fault locations} using the following prompt template ({Prompt Template (Locations)}, abbrev, \texttt{PT-Loc}). Then, if there still are unrepaired faults, we further provide \textbf{the locations of the remaining faults that were not correctly repaired} in the last round using the following prompt template ({Prompt Template (IncR)}, abbrev, \texttt{PT-IncR}):

\begin{figure}[b!]  
    \centering  
    \resizebox{1.0\linewidth}{!}{
        \renewcommand{\arraystretch}{1.1}
        \begin{tabular}{l|c|c}
        \toprule
        \textbf{Prompt Template} & \multicolumn{1}{c|}{\textbf{\# of Corret-Repaired (Y)}} & \textbf{\# of Incorrect-Repaired (N)} \\
        \hline
        PT-B & 16 & 56 \\ 
        \hline
        PT-E & 43 & 29 \\
        \hline
        PT-Loc &  55 & 17\\
        \hline
        PT-IncR &  \textbf{58} & 14\\
        \bottomrule
        \end{tabular}
    }
    \label{fig:table}  
    \resizebox{1.0\linewidth}{!}{
        \includegraphics{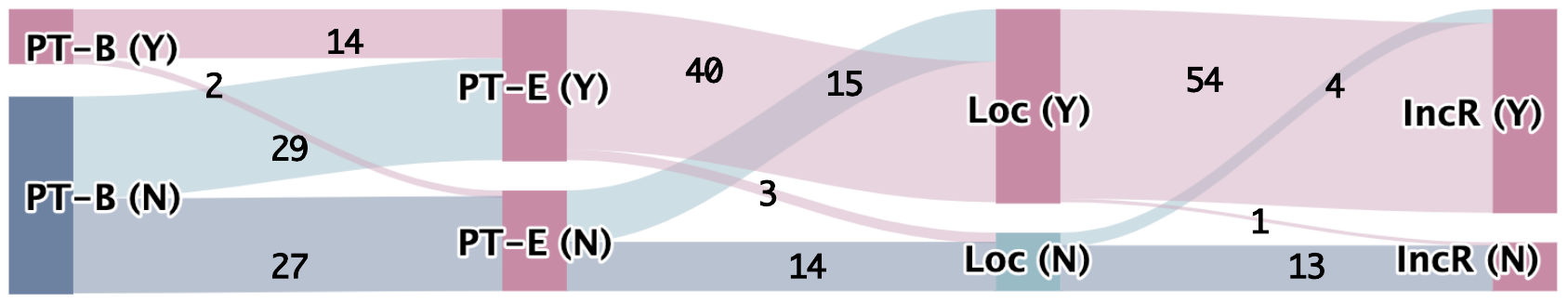}
        }
    \caption{\textbf{The Number of Correctly- (Y) and Incorrectly- (N) Repaired Faults Using Prompt Templates \texttt{PT-B}, \texttt{PT-E} and \texttt{PT-Loc} and \texttt{PT-IncR}.}}
    \label{fig:rq3}
  \end{figure}

\begin{minted}[
    baselinestretch=1.1,
    fontsize=\footnotesize,
    xleftmargin=0.5ex,
    bgcolor=bg,
    breaklines=true,
    escapeinside=||
]{text}
(Round 1) |{\textbf{Prompt Template (Basic)}}|: (Omitted)
(Round 2) |{\textbf{Prompt Template (Enhanced)}}|: (Omitted)
(Round 3) |\underline{\textbf{Prompt Template (Locations)}}| : The repair is wrong. Faults 
may happen in the following locations: |\sethlcolor{highlightY}\hl{[Loc]}|. Please repair it.
(Round 4) |\underline{\textbf{Prompt Template (IncR)}}:| The |\sethlcolor{highlightY}\hl{[Remaining\_loc]}| 
is/are repaired incorrectly. Please repair it accordingly.
\end{minted}

\noindent where the fault location information in \texttt{PT-Loc} is automatically filled in using the information provided by the benchmark~\cite{DeepFD}, while the remaining locations in \texttt{PT-IncR} are filled after analysis.

The results of four rounds are demonstrated in Figure~\ref{fig:rq3}. We visualize the changes in the number of correctly- and incorrectly repaired faults using different prompt templates. Clear that after using the enhanced prompt \texttt{PT-E} (including the information of symptom, learning task, and training dataset), the number of correctly-repaired faults increases from 16 to 43. Then, with the hints of fault locations, the number increases to 55. Finally, with the hints of the unrepaired fault locations, 3 more faults can be repaired. 

In addition, there are a few faults that are repaired correctly in the previous round but changed back to faulty in the next round. This is mainly because ChatGPT does not sure about its repair, so it iteratively repairs these faults without confidence. Besides, we can see that even after we provide the locations of unrepaired faults, there are still 14 faults left unrepaired. 

Taking a closer look, we summarize two reasons for a fault being left unrepaired till the last round of dialogue. 
\textbf{First, ChatGPT ignores the provided information} and repairs other locations instead. An example~\cite{so31627380} is shown below. 

\begin{minted}[
    baselinestretch=1.0,
    fontsize=\scriptsize,
    xleftmargin=0.5ex,
    bgcolor=bg,
    breaklines=true,
    escapeinside=||
]{text}
|\underline{\textbf{Prompt}}:| The repair is wrong. Faults may happen in the following locations:
(1) |\textbf{Activation}|, (2) |\textbf{Optimizer}|. Please repair it again.
|\underline{\textbf{Answer}}:| I apologize for the mistake in my previous response. The issue 
with the code is that |\textbf{the number of units in the output layer}|. 
\end{minted}

\noindent In the prompt, we explicitly provide the fault locations (\ie, activation functions and optimizer), while ChatGPT ignores the hint and locates another fault (\ie, the number of units) instead. This situation may happen due to the catastrophic forgetting~\cite{ramasesh2022effect,korbak2022reinforcement,ramasesh2020anatomy,kemker2018measuring,goodfellow2013empirical,arora2019does}, which is a common problem for the long input sequence.

\textbf{Second, ChatGPT insists the provided location does not faulty because of the misunderstanding of the code intention}. A typical case~\cite{so50306988} is shown below.

\begin{minted}[
    baselinestretch=1.0,
    fontsize=\scriptsize,
    xleftmargin=0.5ex,
    bgcolor=bg,
    breaklines=true,
    escapeinside=||
]{text}
|\underline{\textbf{Prompt}}|:
    |\textbf{num\_classes}| = 2  # Number of output classes
    dummyX, dummyY = |\textbf{make\_multilabel\_classification}|(n_samples = n, 
                        n_features = m, n_classes = |\textbf{num\_classes}|)
    ...
    model.add(keras.layers.Dense(num_classes))  
    model.add(Activation('softmax'))
    model.compile(loss='|\textbf{categorical\_crossentropy}|', 
                                    optimizer='rmsprop')
|\underline{\textbf{Answer}}:| Here's the updated code with the corrected activation and 
loss functions: (No changes are made)
\end{minted}

\noindent The code utilizes the API \texttt{make\_multilabel\_classification} provided by library \texttt{scikit-learn}~\cite{sklearn} to generate the data. While since the number of classes is set to 2, meaning that the training data is generated for a binary classification instead of a multilabel classification problem. Yet, ChatGPT insists that it is a multilabel classification problem, so the inappropriate loss and activation functions remain unrepaired.

\begin{mdframed}[style=MyFrame]
    \textbf{Finding 8}: 
    {\fontsize{10pt}{0}\selectfont\faThumbsOUp}
    With the help of dialogue, 42 (58 - 16) more faults can be correctly repaired.  
    {\fontsize{10pt}{0}\selectfont\faThumbsODown}
    While there are faults that cannot be repaired even with the locations of faults provided due to \textit{\textbf{catastrophic forgetting or the misunderstanding of the code intention}}.
    \end{mdframed}


\section{Case Study}\label{sec:casestudy}
In this section, we summarize the features of ChatGPT 
using examples, and point out possible usage of ChatGPT to facilitate software development. Note that the queries and answers in the following cases are streamlined due to the page limit. 
Only the lines of codes/texts that are of interest are displayed.

\subsection{Code Smell and Refactoring}
Code smells are believed to be implications on software quality and have an impact on maintenance and evolution~\cite{codesmell}. From the analysis of ChatGPT's answers, we notice that ChatGPT is good at recognizing ``bad code smells'' (\eg, dead code, redundant imports), and tends to eliminate them by applying refactoring techniques (\eg, removing dead code, adding code comments). In the following, we use real answers from ChatGPT to demonstrate how it makes suggestions for code smells and code refactoring.

\subsubsection{\textbf{\textit{Naming Convention}}}

There are several smells regarding naming convention~\cite{wake2004refactoring}, such as uncommunicative names and inconsistent names. Though there is no universal standard for the naming convention, it is commonly believed that a desired naming convention expects short but meaningful variable/class names. In our collected answers, two types of smells about the naming convention are identified, but no repaired source codes are provided.

\textbf{Inconsistent Naming Style} occurs if there are more than one naming style in one piece of code. For example, in the following code snippet~\cite{so4544284},
variables are named using either uppercases with underscores 
(\eg, \texttt{X}, \texttt{Y\_encoded}), or lowercases (\eg, \texttt{scaler}). 
ChatGPT suggests to use consistent naming convention (either use lowercase with underscores or camel cases) for variable names.

\begin{minted}[
    baselinestretch=1.0,
    fontsize=\scriptsize,
    xleftmargin=1.5ex,
    bgcolor=bg,
    breaklines=true,
    escapeinside=||
]{python}
|\underline{\textbf{Query}}|: 
    |{\textbf{scaler}}| = StandardScaler()
    |{\textbf{X}}| = scaler.fit_transform(X.reshape(-1, 1))
    |{\textbf{encoder}}| = LabelEncoder()
    |{\textbf{Y\_encoder}}| = encoder.fit_transform(Y)

|\underline{\textbf{Answer}}:| It is recommended to use a consistent naming convention for 
variable names throughout the program. For example, use lowercase with 
underscores or camelCase consistently for all variable names.
\end{minted}

\textbf{Uncommunicative Names}~\cite{wake2004refactoring}
refers to the names used that do not express their intentions adequately, such as \texttt{x} or \texttt{value}. 
For example, in the following code snippet~\cite{so45442843},
ChatGPT points out that the variable names could be more meaningful. It is noteworthy that ChatGPT suggests the variable name to be \texttt{X\_train}, which may indicate that ChatGPT could guess the intention of the variable \texttt{X} to be the definition of the training data, so that it makes such suggestion.

\begin{minted}[
    baselinestretch=1.0,
    fontsize=\scriptsize,
    xleftmargin=1.5ex,
    bgcolor=bg,
    breaklines=true,
    escapeinside=||
]{python}
|\underline{\textbf{Query}}|: 
    X = dataset.values[:, 0].astype(float)
    Y = dataset.values[:, 1]
|\underline{\textbf{Answer}}:| Use more meaningful variable names instead of single letters, e.g.,
"X_train" instead of "X".
\end{minted}

\subsubsection{\textbf{Comments}}
Good code comments are desired to improve code readability and reduce maintenance effort, while irrelevant, trivial or excessive comments may cause confusion to code understanding. During program repair, ChatGPT usually makes suggestions on code comments by adding or deleting them to eliminate smells regarding code comments.

\textbf{Irrelevant Comments} may confuse code understanding and are better removed. In the following code snippet~\cite{so48594888},
a comment \texttt{\#fix1} is placed before the method call \texttt{model.fit}. The comment is irrelevant to the function. ChatGPT identifies the mismatching between the comment and code intention, assuming that this comment is a leftover from a code change and suggesting removing it. 
{This case also indicates that ChatGPT could match the comment semantics to code semantics.}

\begin{minted}[
    baselinestretch=1.0,
    fontsize=\scriptsize,
    xleftmargin=1.5ex,
    bgcolor=bg,
    breaklines=true,
    escapeinside=||
]{python}
|\underline{\textbf{Query}}|:
    # fix1
    model.fit(train_images[:5000], train_labels[:5000], batch_size=1000)
|\underline{\textbf{Answer}}|: The program seems to be correct. The only thing that can be 
improved is the comment before "model.fit()" where it says "fix1". 
This seems to be a leftover from a previous edit and can be removed.
\end{minted}

\textbf{Missing/Inadequate Comments} may hurt code readability and cost extra time for maintainers to read and understand the code functionality from a pile of codes before refactoring. ChatGPT is equipped with code understanding ability, from which the smell of inadequate comments could be eliminated/alleviated. For example, in the following code snippet~\cite{so44998910},
ChatGPT separates the source code into several independent functionality blocks (\eg, data generation, model construction, model training), and adds comments to improve the code readability.

\begin{minted}[
    baselinestretch=1.0,
    fontsize=\scriptsize,
    xleftmargin=0.5ex,
    bgcolor=bg,
    breaklines=true,
    escapeinside=||
]{python}
|\underline{\textbf{Answer}}|: The program could benefit from adding comments to explain the code.
Here's the repaired code with comments added:
    # generate training data                        |\textbf{$\triangleleft$ Added by ChatGPT}|
    x_train = np.random.rand(9000)
    y_train = x_train**4 + x_train**3 - x_train
    x_train = x_train.reshape(len(x_train), 1)
    ...... (omitted due to space limitation)
    
    # train the model                               |\textbf{$\triangleleft$ Added by ChatGPT}|
    model.fit(x_train, y_train, epochs=20, batch_size=50, verbose=1)

    # make predictions on the testing data          |\textbf{$\triangleleft$ Added by ChatGPT}|
    predictions = model.predict(x_test, batch_size=1)
\end{minted}

\subsubsection{\textbf{Dead Code}} Dead code is characterized by a variable, attribute, or code fragment that is not used anywhere~\cite{wake2004refactoring}. It is usually a result of a code change with improper cleaning. For example, it is common that developers import redundant libraries. It may cause dependency conflicts and take up unnecessary resources. ChatGPT could help to identify the redundant imports and remove them correspondingly. For example, in the below code snippet~\cite{so48221692},
ChatGPT removes the unused and duplicated imports, and also abbreviates \texttt{numpy} as \texttt{np} for simplicity.

\begin{minted}[
    baselinestretch=1.0,
    fontsize=\scriptsize,
    xleftmargin=0.5ex,
    bgcolor=bg,
    breaklines=true,
    escapeinside=||
]{python}
|\underline{\textbf{Query}}|: 
    import numpy
    import matplotlib.pyplot as plt          |\textbf{$\triangleleft$ Unused imports}|
    import keras                             |\textbf{$\triangleleft$ Duplicated imports}|
    |\underline{numpy}|.random.seed(seed)

|\underline{\textbf{Answer}}|: The program is not buggy, but it could be improved by removing 
the commented-out code and unused imports. Here's a cleaned-up version:
    import numpy |\underline{as np}|                        |\textbf{$\triangleleft$ Abbreviated by ChatGPT}|
    ...
    |\underline{np}|.random.seed(seed)
\end{minted}

\begin{mdframed}[style=MyFrame]
    \textbf{Finding 9:} 
    {\fontsize{8pt}{0}\selectfont\faThumbsOUp ~\faStar}
    ChatGPT not only understands the semantics of the code snippets, but also understands the comments. By paralleling both semantics, ChatGPT could also detect bad code smells on comments.
\end{mdframed}

\subsection{API Usage}
DL programs are usually built upon DL libraries such as Keras, Tensorflow, and PyTorch. Properly using APIs provided by these libraries could sometimes be tricky even for experienced developers. Regarding API usage, several major issues could be identified and repaired by ChatGPT. We discuss several representative cases in the following.

\subsubsection{\textbf{API Evolution}} 
DL libraries evolve rapidly. The fast API updates bring about both advantages and disadvantages to agile development. Based on our observation, ChatGPT is helpful for warning and repairing issues caused by API evolution. In the following, we discuss two situations.

\textbf{API Deprecation} is a common progression in API lifecycle. With newer version of library released, an API may be outdated and removed from newer version, or whose function is integrated into other APIs. ChatGPT handles API deprecation well by identifying deprecated APIs and recommending new APIs that satisfy the desired functionality. For example, for the following code~\cite{so50079585},
the code under repair calls \texttt{Model.fit\_generator} which takes a data generator as input. While as stated in Tensorflow v2.1 documentation\footnote{\url{https://www.tensorflow.org/versions/r2.1/api\_docs/python/tf/keras/Model\#fit}}, this API is deprecated, and its functionality is integrated into \texttt{Model.fit}. Note that since ChatGPT is trained on the data collected till September 2021, its knowledge about library documentations are up till then. It means that the repair could be invalid once newer version is released.

\begin{minted}[
    baselinestretch=1.0,
    fontsize=\scriptsize,
    xleftmargin=0.5ex,
    bgcolor=bg,
    breaklines=true,
    escapeinside=||
]{python}
|\underline{\textbf{Query}}|:
    model.|\sethlcolor{highlightY}\hl{fit\_generator}|(train_generator) # simplified due to space limitation
|\underline{\textbf{Answer}}|: The program looks fine. However, depending on the version of Keras 
and TensorFlow used, there might be some deprecation warnings related to the 
|\sethlcolor{highlightY}\hl{fit\_generator}| function. In that case, it is recommended to use the |\sethlcolor{highlightY}\hl{fit}|
function instead. 
\end{minted}

\textbf{Parameter Change} is also commonly-seen in software evolution. 
For example, from  Keras v2.0, the parameter \texttt{nb\_epoch} has been renamed to \texttt{epochs}. In the following case~\cite{so31627380}
ChatGPT repairs the outdated parameter to the newer one.

\begin{minted}[
    baselinestretch=1.0,
    fontsize=\scriptsize,
    xleftmargin=0.5ex,
    bgcolor=bg,
    breaklines=true,
    escapeinside=||
]{python}
|\underline{\textbf{Query}}|:
    model.fit(X_train, y_train, |\sethlcolor{highlightY}\hl{nb\_epochs}|=epochs, batch_size=batch_size)
|\underline{\textbf{Answer}}|: The |\sethlcolor{highlightY}\hl{nb\_epochs}| parameter in the "model.fit()" function has been 
deprecated in newer versions of Keras. It is better to use the |\sethlcolor{highlightY}\hl{epochs}| 
parameter instead.
\end{minted}

However, we observe that in some cases, the parameter change could affect the ChatGPT's detection, causing false alarms of repair.
there are cases where the parameters changes escape from ChatGPT's detection. 
For example, given the same parameter change as above (\texttt{np\_epochs} $\rightarrow$ \texttt{epochs}), in the following case~\cite{so39525358},
the change of parameter name leads to a failing detection of API exists. So ChatGPT repairs the program by adding an API already called.

\begin{minted}[
    baselinestretch=1.0,
    fontsize=\scriptsize,
    xleftmargin=0.5ex,
    bgcolor=bg,
    breaklines=true,
    escapeinside=||
]{python}
|\underline{\textbf{Query}}|:
    model.fit(X_train, y_train, nb_epoch=150, batch_size=10)
|\underline{\textbf{Answer}}|: |\textbf{This program seems to be missing the call to}| "fit()" |\textbf{function}|. So, 
assuming the data and parameters are correctly defined, a possible fix 
would be to add the missing "fit()" function call:
    model.fit(X_train, y_train, epochs=150, batch_size=10)
\end{minted}

\begin{mdframed}[style=MyFrame]
    \textbf{Finding 10:} 
    {\fontsize{10pt}{0}\selectfont\faThumbsOUp ~\faStar}
    ChatGPT is helpful for \textbf{detecting the deprecated APIs} regarding the latest libraries it was trained on. It may be used to \textbf{facilitate the software evolution}. 
    {\fontsize{10pt}{0}\selectfont\faThumbsODown}
    However, \textbf{ChatGPT fails to detect a deprecated API even} \textbf{though the change is small in some cases}, which may pose a challenge to using ChatGPT for software evolution.
\end{mdframed}

\subsubsection{\textbf{API Misuse}} API misuse, also known as incorrect API usage, are violated usage of constraints, or non-optimal usage of APIs~\cite{APImisuse}. We observe that ChatGPT is able to repair API misuse by code refactoring, or making suggestions if ChatGPT is not sure about the code intention.

\textbf{API Optimization} improves programs by using APIs more appropriately. For example, for the following code snippet~\cite{so45442843},
ChatGPT optimizes the usage of API \texttt{model.add} by (1) \textbf{Replacing unnamed parameter passing with named parameter passing} to improve the code readability. (2) \textbf{Using parameter in one API instead of using two separate ones} for code simplicity. 

\begin{minted}[
    baselinestretch=1.0,
    fontsize=\scriptsize,
    xleftmargin=0.5ex,
    bgcolor=bg,
    breaklines=true,
    escapeinside=||
]{python}
|\underline{\textbf{Query}}|:
    model.add(Dense(1, input_dim=1))
    model.add(Activation('sigmoid'))
|\underline{\textbf{Answer}}|: ... 3. It is better to use "units" instead of "1" and "2" for the 
"Dense" layers for clarity and readability. 4. The "activation" parameter 
should be used instead of "Activation" layers for simplicity.
    model.add(Dense(|\sethlcolor{highlightY}\hl{units}|=1, input_dim=1, |\sethlcolor{highlightY}\hl{activation}|='sigmoid'))
\end{minted}

\textbf{API Recommendation} usually involves code understanding. It requires certain knowledge of the code intention, then suggestions are made toward the desired intention. We observe several cases where ChatGPT suggests adding APIs to achieve code intention based on its understanding. 
For example, in the following case~\cite{so50306988},
ChatGPT suggests adding an evaluation step after model training.

\begin{minted}[
    baselinestretch=1.0,
    fontsize=\scriptsize,
    xleftmargin=0.5ex,
    bgcolor=bg,
    breaklines=true,
    escapeinside=||
]{python}
|\underline{\textbf{Query}}|:
    model.fit(x_train, y_train, epochs=5, batch_size=32, verbose=True)
|\underline{\textbf{Answer}}|: I would recommend adding a final evaluation step to compute the 
accuracy of the trained model on the test set, which can be done using the 
"model.evaluate()" method.
\end{minted}

Similarly, a recommendation could also be made about the parameter values. For the following example~\cite{so45442843},
the parameter \texttt{verbose} is set to be \texttt{2}.
While ChatGPT suggests setting the parameter \texttt{verbose} to be \texttt{1} to get a progress bar for better visualization during training.

\begin{minted}[
    baselinestretch=1.0,
    fontsize=\scriptsize,
    xleftmargin=0.5ex,
    bgcolor=bg,
    breaklines=true,
    escapeinside=||
]{python}
|\underline{\textbf{Query}}|:
    model.fit(X_train, Y_train, epochs=10, batch_size=200, verbose=2)
|\underline{\textbf{Answer}}|: The "verbose" parameter should be set to "1" to get progress updates 
during training.
\end{minted}

Besides API and parameter recommendation, ChatGPT also suggests the way to use APIs. For example~\cite{so55328966},
the following code snippet uses \texttt{pandas.DataFrame.loc()} to slice rows by names/labels (\texttt{`1'} to \texttt{`5000'}), while ChatGPT suggests to use another API, \texttt{pandas.DataFrame.iloc()}, which selects rows/columns by Integer index/position. Indeed, both APIs could be used to achieve the intention (data slicing). ChatGPT understands the code intention and provides an alternative API for use reference.

\begin{minted}[
    baselinestretch=1.0,
    fontsize=\scriptsize,
    xleftmargin=0.5ex,
    bgcolor=bg,
    breaklines=true,
    escapeinside=||
]{python}
|\underline{\textbf{Query}}|:
    train = train.|\sethlcolor{highlightY}\hl{loc}|['1':'5000', :]
    test = test.|\sethlcolor{highlightY}\hl{loc}|['1':'2000', :]
|\underline{\textbf{Answer}}|: The program seems to be mostly correct, but there might be an issue 
with the data preprocessing part. The "loc" method is being used with string
arguments to subset the data frames, which is not a valid way to select rows 
and columns. Instead, the "iloc" method should be used with integer arguments 
to select rows and columns by index.
    train = train.|\sethlcolor{highlightY}\hl{iloc}|[1:5000, :]
    test = test.|\sethlcolor{highlightY}\hl{iloc}|[1:2000, :]
\end{minted}

\begin{mdframed}[style=MyFrame]
    \textbf{Finding 11:} 
    {\fontsize{10pt}{0}\selectfont\faThumbsOUp ~\faStar}
    The detection of API misuse and API recommendation reflects ChatGPT's code understanding ability. These abilities could facilitate program debugging and improvement.
\end{mdframed}

\section{Discussion}

\subsection{Automated Prompt Template Filling}
Apart from the user scenarios where the templates are filled by developers, we also discuss a possible solution to automate the prompt template filling. Suppose the question description is available, while we want to automatically fill in the template, it is possible to try out the natural language understanding ability of ChatGPT. 
We propose a few-shot prompt template as follows:

\begin{minted}[
    baselinestretch=1.0,
    fontsize=\scriptsize,
    xleftmargin=0.5ex,
    bgcolor=bg,
    breaklines=true,
    escapeinside=||
]{text}
|\underline{\textbf{Description}}|: [TEXT]
|\underline{\textbf{Template}}|  : The following code is designed for a [X] problem 
trained on [Y] dataset. Please repair it in order to [Z].
|\underline{\textbf{Example}}|   : The following code is designed for a [classification] 
problem trained on [Iris] dataset. Please repair it in order to 
[improve the accuracy].
|\textbf{Please fill in the [X], [Y], [Z] in the Template according to the}| 
|\textbf{Description, and output the filled template only.}|
\end{minted}

\noindent The answer from ChatGPT is shown as follow (Answer). Compared with the human-written prompt (Reference), the meaning of two sentences are almost identical, which indicates the possibility to automate the prompt as discussed in RQ2. 

\begin{minted}[
    baselinestretch=1.0,
    fontsize=\footnotesize,
    xleftmargin=0.5ex,
    bgcolor=bg,
    breaklines=true,
    escapeinside=||
]{text}
|\underline{\textbf{Answer}}:|   The following code is designed for a |\textbf{classification}| 
problem trained on the |\textbf{MNIST}| dataset. Please repair it in order to 
|\underline{\textbf{improve the network's learning}}|.
|\underline{\textbf{Reference}}:| The following code is designed for a |\textbf{classification}| 
problem trained on the |\textbf{MNIST}| dataset. Please repair it in order to 
|\underline{\textbf{get a higher accuracy}}|.
\end{minted}

\subsection{Failing-Test-Guided Prompt}
A recent study~\cite{nashidretrieval} utilizes the failing test as hints to improve the repair ability. {However, this study only focuses on conventional programs while the decision logic of DL programs is not directly encoded in the source code.}  We are thus motivated to explore whether failing tests could also help DL program repair.
We take an instance from StackOverflow~\cite{so31556268}
as an example. This case contains four faults in the program, while using all the above prompts (PT-B, PT-E), there are at most 1 fault that can be correctly repaired. After providing the hint of fault locations, there is still one fault unrepaired. We then try whether the information of failing tests could complete the repair.

In particular, this case trains a DNN to emulate the XOR problem, 
we first train the repaired program returned by ChatGPT. With the hints of fault locations, ChatGPT successfully repaired three faults. We execute the program and test the trained DNN against the test set. A failing test appears, \ie, given the input \texttt{[1,1]}, the DNN outputs \texttt{[9.9999845e-01]} ($\approx$ \texttt{[1]}), while the expected output is \texttt{[0]}. We then integrate such failing information to ChatGPT by saying ``\textit{The repair is wrong. For the input \texttt{[1, 1]}, the output should be \texttt{[0]} instead of \texttt{[1]}. Please repair it again.}'' ChatGPT apologizes and then adds a rounding step before the output. We continue asking it to output a number that is \textit{close} to \texttt{[1]}, but ChatGPT still cannot repair the fault. It indicates that for DL program repair, providing the failing test cases may not be helpful.

\begin{mdframed}[style=MyFrame]
    \textbf{Finding 12:} 
    {\fontsize{10pt}{0}\selectfont\faThumbsODown}
    Providing \textit{\textbf{an individual failing test to ChatGPT may not help DL program repair}} because the DL programs do not explicitly write the decision logic of the resulting deep neural networks.
    Besides, ChatGPT is not able to execute the program at the current stage, even though a patch is provided by ChatGPT, a validation step is still needed.
\end{mdframed}

\section{Related Work}

\subsection{Large Language Model for Software Engineering (LLM4SE)}
With the breakthrough of LLMs, studies have proposed to explore how LLMs can be used to assist developers in a variety of tasks, such as code generation~\cite{codeUnderstanding,poesia2022synchromesh}, program repair~\cite{sobania2023analysis,llmEra,codexLLM,jiang2023impact,nashidretrieval,refactoring,pearce2021can}, and code summarization~\cite{codeUnderstanding,fewshot-codesummary}. 
In particular, Zeng \etal~\cite{codeUnderstanding} conducted an extensive study, evaluating eight pre-trained LLMs and four non-pre-trained models against three code understanding tasks (\ie, defect detection, clone detection, code search) and four code generation tasks (\ie, code summarization, code repair, code translation, code generation). Their evaluation result indicates that pre-trained LLMs are more promising than non-pre-trained models. 
Recently, with the advent of ChatGPT {and InstructGPT~\cite{trainingHF} fine-tuned using RLHF~\cite{RLHF}}, researchers devoted to exploring ChatGPT's abilities in these major code-related tasks such as code repair~\cite{sobania2023analysis} and code generation~\cite{gozalo2023chatgpt}, as well as in other related research areas such as natural language processing~\cite{bang2023multitask}, mathematics~\cite{gilson2022well,frieder2023mathematical,davis2023mathematics}, reasoning~\cite{guo2023close,davis2023mathematics}. 
These studies show the potential of ChatGPT in these areas, point out the rooms to improve, and shed some light on how to use ChatGPT properly to facilitate the corresponding tasks.
In particular, Sobania \etal~\cite{sobania2023analysis} evaluate ChatGPT on conventional Python programs with basic functionalities (\eg, sorting, searching), examining the program repair ability.
Our work complements their study of using buggy programs with more complicated functionalities and more dependencies. Besides, our study also compares the results using a different prompt template and summarizes the pros and cons of ChatGPT's reply.

\subsection{Deep Learning Program Repair}
To debug and repair DL programs, various approaches have been proposed.
AutoTrainer~\cite{autotrainer} proposed an automatic approach to detect and fix the training problems in DNN programs at runtime. It particularly focuses on detecting and repairing five common training problems: gradient vanish, gradient explode, dying ReLU, oscillating loss, and slow convergence. It encapsulates and automates the detecting and repairing process by dynamic monitoring. Amazon SageMaker Debugger~\cite{rauschmayr2021amazon} and UMLAUT~\cite{Umlaut21} provide a set of built-in heuristics to debug faults in DNN models during training. 
DeepLocalize~~\cite{wardat21DeepLocalize}, the first fault localization approach for DNN models (such as incorrect learning rate or inappropriate loss function). It locates in layers where the symptom happens instead of where the fault's root cause resides. DeepFD~\cite{DeepFD} and DeepDiagnosis~\cite{DeepDiagnosis} diagnose and localize faults in DL programs. DeepDiagnosis also suggests actionable advice on how to repair. 

\section{Threats of Validity}
To avoid the difference between the ChatGPT website interface and the latest API, we repeat the experiment of RQ1 using the interface (Mar 14, 2023 Version). 
The experiment results only show a slight difference. The evaluation results are released for the reference. Besides, ChatGPT's replies are carried out manually, which may be subjective and misunderstandings. To alleviate this threat, the cases used in the paper come along with StackOverflow post IDs. Also, all the prompts and replies could be traced in our released repository. 
{Finally, since ChatGPT is trained on a large corpus of text data from the Internet (\eg, books, articles, open-source code repositories), it is possible that there is a data leakage problem, \ie, the cases in the benchmark were included in its training set, so the output is generated due to memorization. 
However, we observe that ChatGPT does not perform well with various prompt templates. It indicates it has not memorized the cases in the benchmark. This observation may partially alleviate the concern regarding such threat.}
\section{Conclusion}
In this paper, we study ChatGPT's ability in DL program debugging (\ie, fault detection, fault localization, and program repair). In particular, we explore the impact of prompts on the debugging performance and propose a prompt template to improve the performance. In addition, we explored how far ChatGPT's dialogue feature can push the performance forward. On top of the experiment, we summarize the pros and cons of ChatGPT's reply and point out the possible directions where ChatGPT could help to facilitate the SE community.
\section{Statement of AI Tool Usage.}
During the paper writing, we use Grammarly~\cite{grammarly} to check grammar and use ChatGPT to rephrase the title of the paper, as well as to conduct the experiment.



\bibliographystyle{ACM-Reference-Format}
\bibliography{reference}


\end{document}